\documentstyle[11pt,epsf,fleqn]{article}
\begin{document}
\title{Electroweak phase transition in the standard model with a dimension-six  Higgs operator at one-loop level}
\author{S. W. Ham$^{(1)}$ and S. K. Oh$^{(1,2)}$
\\
\\
{\it $^{(1)}$ Center for High Energy Physics, Kyungpook National University}\\
{\it Daegu 702-701, Korea} \\
{\it $^{(2)}$ Department of Physics, Konkuk University, Seoul 143-701, Korea}
\\
\\
}
\date{}
\maketitle
\begin{abstract}
The possibility of a strongly first-order electroweak phase transition by means of a dimension-six Higgs operator
in the Higgs potential of the standard model is studied at finite temperature at the one-loop level.
Exact calculation of the one-loop effective Higgs potential at finite temperature suggests that
for the Higgs boson with its mass between 115 and 132 GeV  the strongly first-order electroweak phase transition
is possible if a dimension-six operator is present.
\end{abstract}
\vfil
\eject

\section{Introduction}

The baryogenesis via electroweak phase transition is an interesting scenario to explain the
baryon asymmetry of the universe because it can be tested in the future high energy
experiments [1-3].
Sakharov pointed out several decades ago three ingredients for generating dynamically
the baryon asymmetry of the universe from a baryon-symmetric universe [4].
Among them is the requirement that the universe should depart from the thermal equilibrium.
The departure from thermal equilibrium is required to guarantee the first order
phase transition at the electroweak scale.
The strength of the first order phase transition must be strong enough for preserving the generated baryon asymmetry at the electroweak scale.

Previous calculations indicate that there could be a strongly first-order electroweak phase transition
in the standard model (SM) if the Higgs boson has a mass of less than 35 GeV,
which is well below the experimental lower bound [5-7].
Lattice simulations have shown that the electroweak phase transition actually vanishes completely if $m_H >$ 75 GeV [8].
Thus, the SM in its conventional form eliminates the possibility of a strongly first-order electroweak phase transition.
It seems that the SM should be enlarged in order to revive the possibility.
A number of suggestions have been published in the literature within the framework of the SM for the strongly first-order electroweak phase transition [9-17].

One line of thoughts is based on high dimensional operators.
In the SM, the phenomenological implications of the high-dimensional operators are
considered in various papers [18-21].
It is already investigated that a strongly first order phase transition
in the context of the electroweak phase transition is allowed
for a relatively large Higgs boson mass ($\sim$ 100 GeV) by considering
a dimension-six Higgs operator relevant to the tree-level Higgs potential of the SM [22].
Recently, the electroweak phase transition in the SM is investigated
by considering a low cutoff on the dimension-six Higgs operator [23].

In this paper, we would like to examine the possibility of a strongly first-order electroweak phase transition
in the presence of a dimension-six operator in the Higgs potential, at the one-loop level.
We include the loops of gauge bosons, top quark, Higgs and Goldstone bosons into the one-loop contributions.
We calculate the finite-temperature effective potential at the one-loop level by exact integration.
We search the parameter space where the Higgs boson has a reasonable mass and the where the
electroweak phase transition is both first order and strong enough.
The result of our calculations suggests that a strongly first-order electroweak phase transition is possible in the SM
with a dimension-six operator if the Higgs boson has a mass between 115 GeV and 132 GeV.

\section{The Electroweak Phase Transition}

The Higgs potential in the SM consists of one Higgs doublet
\begin{eqnarray}
\Phi & = & {1 \over \sqrt{2}}
  \left ( \begin{array}{c}
          \phi^+  \cr
          \phi + i \eta
  \end{array} \right )    \ ,
\end{eqnarray}
where the real component of the neutral Higgs field may be expressed as $\phi = v + H$,
where $v$ is the vacuum expectation value (VEV) for the electroweak symmetry breaking
and $H$ is the physical Higgs boson.
The conventional form of the tree-level Higgs potential in the SM up to $\phi^4$, at zero temperature, is
\begin{equation}
    V_0(\phi, 0) =  - {\mu^2 \over 2} \phi^2 + {\lambda \over 4} \phi^4 \ .
\end{equation}
To this Higgs potential, we introduce a non-renormalizable dimension-six term, which may be defined as
\begin{equation}
    O_6  = {\alpha \over 8 \Lambda^2} (\phi^2-v^2)^3 \ ,
\end{equation}
where $\alpha$ is a parameter for weakly coupled new physics and $\Lambda$ is a cutoff.

The one-loop contribution at zero temperature is given by the effective potential method as [24,9,10]
\begin{equation}
    V_1(\phi, 0)
    =  \sum_i {n_i \over 64 \pi^2}
    \left [ m_i^4 (\phi) \log \left ({m_i^2 (\phi) \over m_i^2 (v) } \right )
    - {3 \over 2} m_i^4 (\phi) + 2 m_i^2 (v) m_i^2 (\phi) \right ] \ ,
\end{equation}
where $i = W$, $Z$, $t$, $\phi$, and $G$, respectively for $W$ boson, $Z$ boson, top quark,
the Higgs boson, and the Goldstone boson.
The degrees of freedom for each particle are $n_W = 6$, $n_Z = 3$, $n_t = - 12$,
$n_{\phi} = 1$, and $n_G = 3$, and the field-dependent masses for each particle are given by
$m_W^2(\phi) = g_2^2 \phi^2/4 $,
$m_Z^2(\phi) = (g_1^2 + g_2^2) \phi^2/4$,
$m_t^2(\phi) = h_t^2 \phi^2/2$,
$m_{\phi}^2(\phi) = 3 \lambda \phi^2 - (9 v^2  / 2 \Lambda^2) \phi^2$,
and $m_G^2(\phi) = \lambda \phi^2 - (3 v^2  /2 \Lambda^2) \phi^2$.
Thus, the effective potential at the one-loop level for our purpose is
\begin{equation}
    V(\phi, 0) = V_0 (\phi, 0) + O_6 + V_1 (\phi, 0)
\end{equation}
at zero temperature.

The mass of the physical Higgs boson, $m_H$, is obtained by substituting $\phi = v$
in the second derivative of $V(\phi, 0)$.
Note that both $m_{\phi}^2$ and $m_G^2$ should be positive in order to be employed in $V_1(\phi, 0)$.
The condition that they should be positive yields a cutoff for $\Lambda \ge \sqrt{3} v^2/m_H$.
A typical value of the cutoff for $\Lambda$ is 912 GeV, which is obtained from $v = v_0$
and the experimental lower bound on the Higgs boson mass, $m_H \ge 115$ GeV, for $\alpha = 1$.
Hereafter, we will set $\alpha = 1$ for simplicity, unless explicitly specified otherwise.
One might consider the case of $\alpha \neq 1$ or the case of negative $\alpha$.

In Fig. 1a, we plot the Higgs potential at zero temperature as a function of $\phi$
for $\alpha = 1$, $\Lambda = 912$ GeV, $m_H$ = 115 GeV.
There are four curves in the figure:
The solid curve represents $V(\phi, 0) = V_0(\phi,0) + V_1 (\phi, 0)$, and the dashed curve
$V(\phi, 0) = V_0(\phi,0) + V_1 (\phi, 0) + O_6$.
In the remaining two curves, we neglect Higgs and Goldstone boson loops in the one-loop contribution.
Denoting the one-loop contribution without Higgs and Goldstone boson loops as $V_{1-(\phi,G)}(\phi, 0)$,
we plot $V(\phi, 0) = V_0(\phi,0) + V_{1-(\phi,G)}(\phi, 0)$ (dotted curve) and
$V(\phi, 0) = V_0(\phi,0) + V_{1-(\phi,G)} (\phi, 0) + O_6$ (dot-dashed curve).

One can see that the solid curve is nearly identical to the dotted one and the dashed curve is
almost identical to the dot-dashed one.
This implies that the contributions of Higgs and Goldstone bosons in the one-loop effective potential is
negligible, for $\alpha = 1$, $\Lambda = 912$ GeV, $m_H$ = 115 GeV.
On the other hand, the effect of the presence or the absence of the dimension-six term, $O_6$, is significant,
as the two groups of curves show distinctively different patterns.
However, all of the the four curves show a minimum at $\phi = v = v_0 = 246$ GeV, for finite $\phi$.
The VEV $v$ of $\phi$ is defined as the minimum of the potential.
As $\phi$ increases indefinitely, the potential without $O_6$ will eventually decrease to the negative infinity.
Therefore, the minimum at  246 GeV is the global minimum for the potential with $O_6$,
whereas it is a local minimum for the potential without $O_6$.
This is already observed in a number of papers, for example, from Ref. [25].
At the higher-order level the Higgs potential may escape from decreasing indefinitely.

Without the one-loop contribution $V_1 (\phi, 0)$, it has been observed that $\phi = v_0$
may not be the zero-temperature minimum of the potential [23].
In our case, the presence of the $O_6$ term ensures that the minimum of the potential occurs
always at $\phi = v_0$ at zero temperature.
Thus, the minimum at 246 GeV is the global one in our case.

In Fig. 1b, we plot the Higgs potential at zero temperature as a function of $\phi$.
In this figure, we set $\alpha = -1$.
The other parameters are the same as Fig. 1a: $\Lambda = 912$ GeV, $m_H$ = 115 GeV.
There are two curves: The solid curve represents  $V(\phi, 0) = V_0(\phi,0) + V_1 (\phi, 0) + O_6$,
and the dashed curve does $V(\phi, 0) = V_0(\phi,0) + V_{1-(\phi,G)} (\phi, 0) + O_6$.
Here, in Fig. 1b, the contribution of Higgs and Goldstone bosons is clearly significant.
Moreover, the minimum at $\phi = 246$ GeV for the solid curve is not a global one.
Thus, Fig. 1b indicates that the one-loop effective potential with $O_6$ for negative $\alpha$
exhibits the same difficulty as the potential without $O_6$ for positive $\alpha$.
Hereafter, we do not consider the case of negative $\alpha$ in the dimension-six Higgs operator.
A detailed analysis for negative $\alpha$ can be found in Ref. [20].

Now, we study the case of finite temperature.
The one-loop contribution at finite temperature is given by [26]
\begin{equation}
V_1 (\phi, T)  =  \sum_i {n_i T^4 \over 2 \pi^2}
            \int_0^{\infty} dx \ x^2 \
            \log \left [1 \pm \exp{\left ( - \sqrt {x^2+{m_i^2(\phi)/T^2 }} \right )  } \right ] \ ,
\end{equation}
where the negative sign is for bosons and the positive sign for fermions.
Here, we have $i = W$, $Z$, $t$, $\phi$, and $G$, respectively for $W$ boson, $Z$ boson, top quark,
the Higgs boson, and the Goldstone boson, and $n_W = 6$, $n_Z = 3$, $n_t = - 12$, $n_{\phi} = 1$, and $n_G = 3$.
The full one-loop effective potential at finite temperature that we are considering can now be expressed as
\begin{equation}
    V(\phi, T) = V_0(\phi,0) + O_6 + V_1(\phi, 0) + V_1(\phi, T) \ .
\end{equation}

The above one-loop contribution at finite temperature may be approximated for high temperature as [6,9,10]
\begin{eqnarray}
V_1^{({\rm high} \ T)} (\phi, T) & = &\mbox{} - n_t \left [{T^2 m_t^2 (\phi) \over 48}
+ {m_t^4 (\phi) \over 64 \pi^2} \log \left ({m_t^2 (\phi) \over a_f T^2} \right ) \right ] \cr
& &\mbox{} + \sum_i n_i \left [{T^2 m_i^2 (\phi) \over 24} - {T m_i^3 (\phi) \over 12 \pi}
- {m_i^4 (\phi) \over 64 \pi^2} \log \left ({m_i^2 (\phi) \over a_b T^2} \right ) \right ] \ ,
\end{eqnarray}
where $\log a_f = 1.14$ and $\log a_b = 3.91$.
It is known that in the SM the high temperature approximation is consistent
with the exact calculation of the integrals within 5 \% for $m_f/T < 1.6$ in the fermion case
and for $m_b/T < 2.2$ in the boson case.
We would not only use the high temperature approximation but also perform the exact integrations
in $V_1(\phi, T)$ numerically.

The two curves in Figs. 2a and 2b show the results of the high temperature approximation as well as the exact integrations.
In Fig. 2a, we plot $V(\phi, T)$ in the high temperature approximation, as a function of $\phi$
for $\alpha = 1$, $\Lambda = 912$ GeV, $m_H$ = 115 GeV.
The critical temperature is obtained as $T_c = 117.04$ GeV.
One can see that there are two minima in the potential at $\phi = 0$ GeV and $\phi = 129$ GeV.
These two minima consist of degenerate vacua, and allow the first-order phase transition.
The symmetric phase state corresponds to the vacuum at $\phi = 0$ GeV while the broken phase state $\phi = 129$ GeV.
We denote the critical value of the VEV at the broken phase state as $v_c$, that is, $v_c= 129$ GeV in this case.
The ratio of $v_c$ and $T_c$ is a criterion for the strength of the first order electroweak phase transition.
Generally, it is regarded as a strong phase transition if $v_c/T_c \ge 1$ at electroweak scale.
In Fig. 2a, $v_c/T_c$ is 1.102, thus satisfying the criterion for the strongly first-order electroweak phase transition.

In Fig. 2b, we repeat the job using the result of exact integrations, and plot $V(\phi, t)$
as a function of $\phi$ for $\alpha = 1$, $\Lambda = 912$ GeV, $m_H$ = 115 GeV.
The critical temperature in this case is obtained as $T_c =117.14$ GeV, which is
slightly larger than the result in the high temperature approximation.
The critical VEV in this case is, as one can see from Fig. 2b, $v_c =  134$ GeV.
Thus, we have $v_c/T_c = 1.134$ in Fig. 2b.
Both the high temperature approximation and the exact integration suggest that for $m_H = 115$ GeV
there can be a strongly first order phase transition at the electroweak scale,
if the Higgs potential includes a dimension-six operator.

In this way, we search the parameter space to examine if a strongly first order phase transition at the electroweak scale may take place for other set of parameter values.
The ratio of $v_c/T_c$ decreases, or the strength of the phase transition becomes weak, as $\Lambda$ increases, for fixed Higgs boson mass.
Thus, we carry out exact integrations for $V_1(\phi,T)$, varying $\Lambda$ from the cutoff value upward,
for given Higgs boson mass, and examine whether the ratio of $v_c/T_c$ stays above 1 or not.
We then increase the Higgs boson mass step by step and vary $\Lambda$ from the cutoff value upward
and check the value of $v_c/T_c$.
The largest value of $v_c/T_c$ is actually obtained by taking the smallest, or the cutoff,
value of $\Lambda$, for given Higgs boson mass, since $T_c$ increases and $v_c$ decreases as $\Lambda$ increases.
The result of our search is summarized in Table I, where the result is listed
in the increasing order of the Higgs boson mass.
The first set of entries corresponds to the result of Fig. 2b.
Note that the cutoff value decreases as the Higgs boson mass increases, $\Lambda \ge \sqrt{3} v^2/m_H$.

\begin{table}[ht]
\caption{Strongly first order electroweak phase transition for various Higgs boson masses for $\alpha = 1$.}
\begin{center}
\begin{tabular}{c|c|c|c|c|c}
\hline
\hline
set & $m_H$  & $\Lambda$ & $T_c$ & $v_c$ & $v_c/T_c$ \\
\hline
\hline
1 & 115 & 912 & 117.14 & 134 & 1.143  \\
\hline
2 & 115 & 950 & 119.00 & 127 & 1.067  \\
\hline
3 & 120 & 874 & 120.90 & 133 & 1.100  \\
\hline
4 & 125 & 839 & 124.69 & 132 & 1.058 \\
\hline
5 & 130 & 807 & 128.52 & 131 & 1.019 \\
\hline
6 & 131 & 801 & 129.29 & 131 & 1.013 \\
\hline
7 & 132 & 795 & 130.07 & 130 & 0.999 \\
\hline
8 & 135 & 777 & 132.35 & 130 & 0.982 \\
\hline
9 & 140 & 749 & 136.19 & 130 & 0.954 \\
\hline
\hline
\end{tabular}
\end{center}
\end{table}

Notice that the Higgs boson mass in the first and the second set of entries in Table 1 is the same.
Actually, we fix the Higgs boson mass at $m_H = 115$ GeV and increase $\Lambda$ until the ratio of $v_c/T_c$
decreases down to about 1.
As we increase $\Lambda$ still further, the ratio is found to decrease below 1.
We find that $\Lambda$ may be as large as 950 GeV for $m_H = 115$ GeV in order to ensure
the strongly first-order electroweak phase transition.

One can see in Table I that, if the Higgs boson mass is smaller than 132 GeV, the ratio of $v_c/T_c$
can be larger than or comparable to 1 for a certain value of $\Lambda$ above the cutoff value.
Therefore, Table I tells us that  the strongly first-order electroweak phase transition might take place
if $m_H \le 132$ GeV.
A Higgs boson with its mass between 115 GeV and 132 GeV is within the reach of future experiments.
It can be discovered possibly in the Tevatron experiments or in the LHC experiments in the near future.

\section{Conclusions}

The possibility of the strongly first-order electroweak phase transition in the SM is studied
by introducing a dimension-six Higgs operator to the finite temperature Higgs potential at one-loop level.
The one-loop contributions due to gauge bosons, top quark, Higgs and Goldstone bosons are taken into account.
Finite temperature effects are also included by the one-loop finite-temperature correction.
Instead of employing the high temperature approximation, we carry out the exact integrations.
The difference between the two method of calculations is not serious, but noticeable.
For given Higgs boson mass, we calculate the critical temperature ($T_c$) with two degenerate global vacua
and the critical value of the Higgs field VEV ($v_c$) for the Higgs boson mass.
Then, we examine whether $v_c/T_c$ is larger than 1 or not.
In this way, we study if the condition for a strongly first-order electroweak phase transition is satisfied or not.
We find that there are parameter space where $v_c/T_c$ is larger than or comparable to 1 for $115  \le m_H \le 132$ GeV.
Thus, our study suggests that the SM Higgs potential at the one-loop level with a dimension-six operator allows
a strongly first-order electroweak phase transition, for the Higgs boson mass between 115 GeV and 132 GeV.
A Higgs boson with this mass range might easily be searched in the future high energy experiments.

\vskip 0.3 in

\noindent
{\large {\bf Acknowledgments}}
\vskip 0.2 in
\noindent
This work was supported by Korea Research Foundation Grant (2001-050-D00005).

\vskip 0.2 in



{\bf Figure Captions}

\vskip 0.3 in
\noindent
Fig. 1(a) : The effective potentials $V(\phi, 0) = V_0(\phi,0) + V_1 (\phi, 0)$
(solid curve), $V(\phi, 0) = V_0(\phi,0) + V_1 (\phi, 0) + O_6$ (dashed curve),
$V(\phi, 0) = V_0(\phi,0) + V_W (\phi, 0) +  V_Z (\phi, 0) + V_t (\phi, 0)$ (dotted curve),
and $V(\phi, 0) = V_0(\phi,0) + V_W (\phi, 0) +  V_Z (\phi, 0) + V_t (\phi, 0) + O_6$
(dot-dashed curve), as a function of $\phi$, for $m_H = 115$ GeV and $\alpha = 1$,
$\Lambda = 912$ GeV on $O_6$.
The solid curve overlap nearly the dotted one while the dashed curve overlap exactly dot-dashed one.
The minimums of the potential occur at $\phi = v = v_0 = 246$ GeV for four curves.

\vskip 0.3 in
\noindent
Fig. 1(b) : The effective potential with $O_6$ for $\alpha = - 1$, as a function of $\phi$.
Two curves display for $V(\phi, 0) = V_0(\phi,0) + V_1 (\phi, 0) + O_6$ (solid curve)
and $V(\phi, 0) = V_0(\phi,0) + V_W (\phi, 0) +  V_Z (\phi, 0) + V_t (\phi, 0) + O_6$ (dashed curve).

\vskip 0.3 in
\noindent
Fig. 2(a) : $V(\phi, T)$ for a high temperature approximation by using Eq. (8),
as a function of $\phi$ at the critical temperature, $T_c$ = 117.04 GeV.
There are two degenerate vacua when $\phi = 0$ GeV and $\phi = 129$ GeV.

\vskip 0.3 in
\noindent
Fig. 2(b) : $V(\phi, T)$ for an exact integration calculation on the temperature-dependent
effective potential by using Eq. (6), for the parameter setting of Fig. 2a.
\vfil\eject

\setcounter{figure}{0}
\def\figurename{}{}%
\renewcommand\thefigure{Fig. 1a}
\begin{figure}[t]
\epsfxsize=12cm
\hspace*{2.cm}
\epsffile{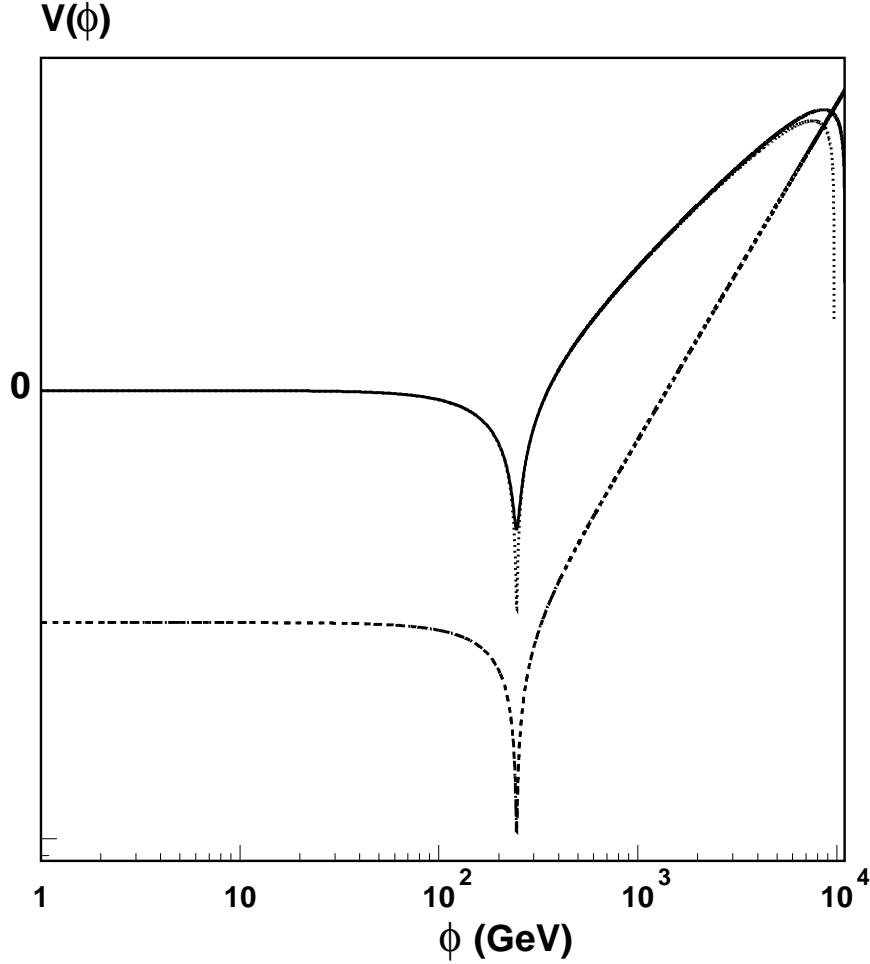}
\caption[plot]{The effective potentials $V(\phi, 0) = V_0(\phi,0) + V_1 (\phi, 0)$
(solid curve), $V(\phi, 0) = V_0(\phi,0) + V_1 (\phi, 0) + O_6$ (dashed curve),
$V(\phi, 0) = V_0(\phi,0) + V_W (\phi, 0) +  V_Z (\phi, 0) + V_t (\phi, 0)$ (dotted curve),
and $V(\phi, 0) = V_0(\phi,0) + V_W (\phi, 0) +  V_Z (\phi, 0) + V_t (\phi, 0) + O_6$
(dot-dashed curve), as a function of $\phi$, for $m_H = 115$ GeV and $\alpha = 1$,
$\Lambda = 912$ GeV on $O_6$.
The solid curve overlap nearly the dotted one while the dashed curve overlap exactly dot-dashed one.
The minimums of the potential occur at $\phi = v = v_0 = 246$ GeV for four curves.}
\end{figure}
\renewcommand\thefigure{Fig. 1b}
\begin{figure}[t]
\epsfxsize=12cm
\hspace*{2.cm}
\epsffile{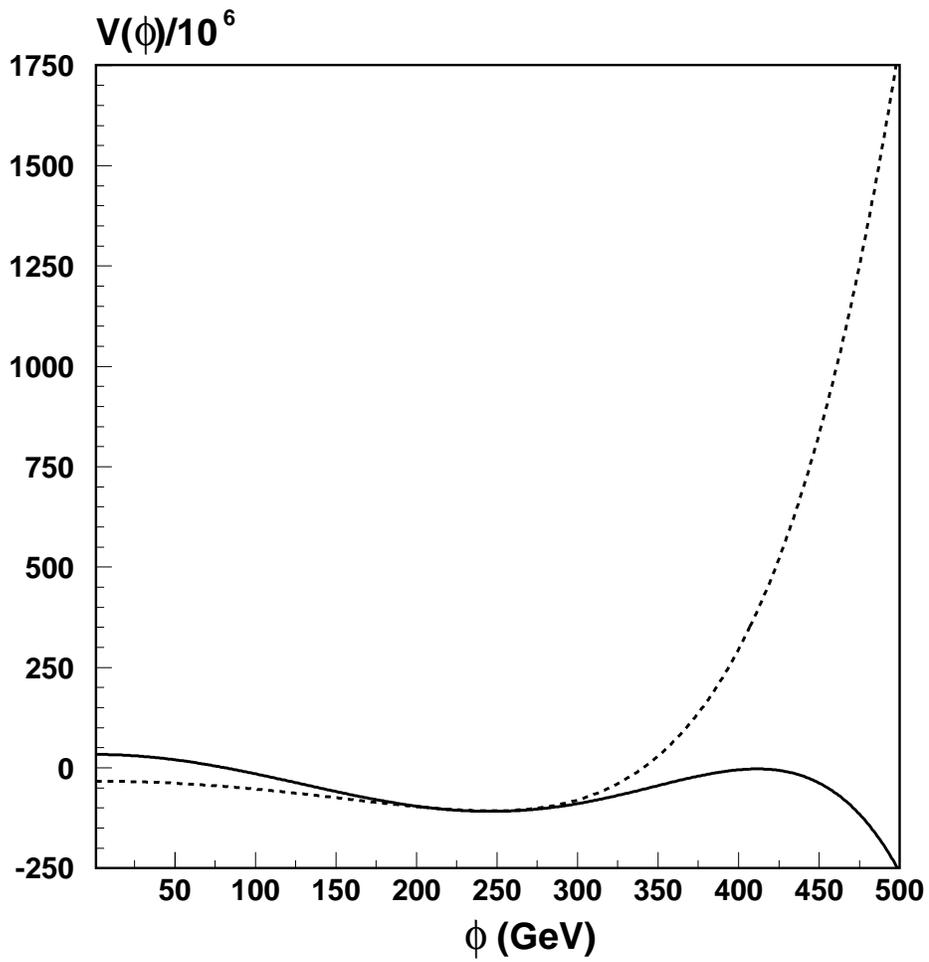}
\caption[plot]{The effective potential with $O_6$ for $\alpha = - 1$, as a function of $\phi$.
Two curves display for $V(\phi, 0) = V_0(\phi,0) + V_1 (\phi, 0) + O_6$ (solid curve)
and $V(\phi, 0) = V_0(\phi,0) + V_W (\phi, 0) +  V_Z (\phi, 0) + V_t (\phi, 0) + O_6$ (dashed curve).}
\end{figure}
\renewcommand\thefigure{Fig. 2a}
\begin{figure}[t]
\epsfxsize=12cm
\hspace*{2.cm}
\epsffile{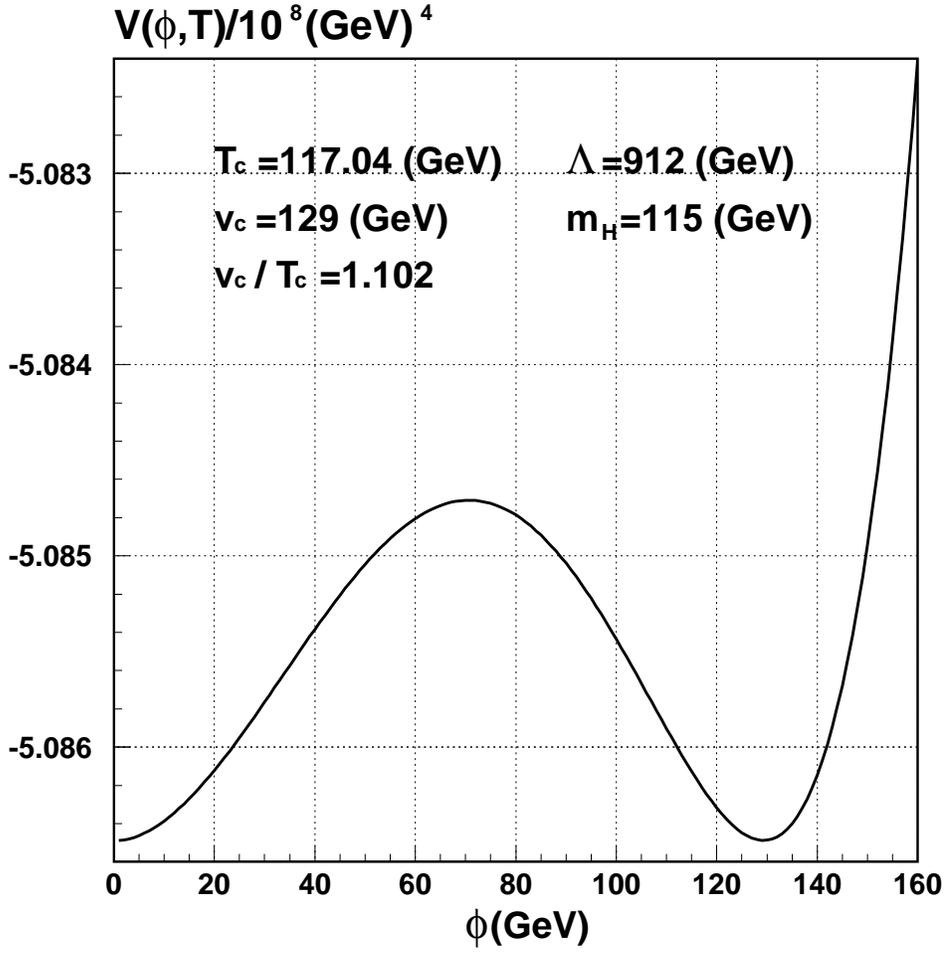}
\caption[plot]{$V(\phi, T)$ for a high temperature approximation by using Eq. (8),
as a function of $\phi$ at the critical temperature, $T_c$ = 117.04 GeV.
There are two degenerate vacua when $\phi = 0$ GeV and $\phi = 129$ GeV.}
\end{figure}
\renewcommand\thefigure{Fig. 2b}
\begin{figure}[t]
\epsfxsize=12cm
\hspace*{2.cm}
\epsffile{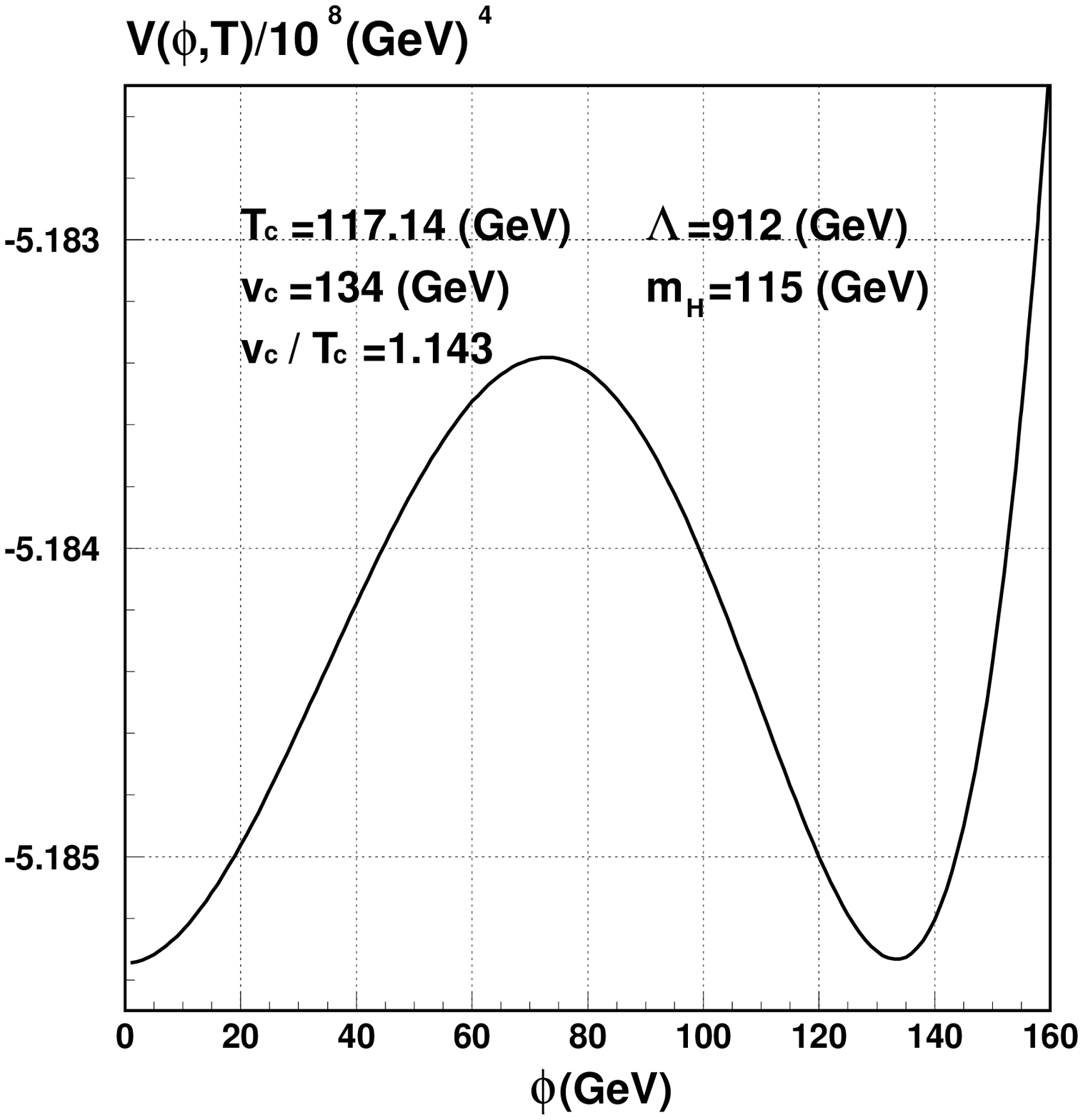}
\caption[plot]{$V(\phi, T)$ for an exact integration calculation on the temperature-dependent
effective potential by using Eq. (6), for the parameter setting of Fig. 2a.}
\end{figure}

\begin{thebibliography}{99}

\bibitem{1}
V.A. Kuzmin, V.A. Rubakov, and M.E. Shaposhnikov, Phys. Lett. B {\bf 155}, 36 (1985).
\bibitem{2}
M.E. Shaposhnikov, JETP Lett. {\bf 44}, 465 (1986);
M.E. Shaposhnikov, Nucl. Phys. B {\bf 287}, 757 (1987);
M.E. Shaposhnikov, Nucl. Phys. B {\bf 299}, 797 (1988);
L.D. McLerran, Phys. Rev. Lett. {\bf 62}, 1075 (1989);
N. Turok and J. Zadrozny, Phys. Rev. Lett. {\bf 65}, 2331 (1990);
N. Turok and J. Zadrozny, Nucl. Phys. B {\bf 358}, 471 (1991);
L.D. McLerran, M.E. Shaposhnikov, N. Turok, and M.B. Voloshin, Phys. Lett. B {\bf 256}, 451 (1991);
M. Dine, P. Huet, R.J. Singleton, and L. Susskind, Phys. Lett. B {\bf 257}, 351 (1991).
\bibitem{3}
A.G. Cohen, D.B. Kaplan, and A.E. Nelson, Annu. Rev. Nucl. Part. Sci. {\bf 43}, 27 (1993);
M. Trodden, Rev. Mod. Phys. {\bf 71}, 1463 (1999);
A. Riotto and M. Trodden, Annu. Rev. Nucl. Part. Sci. {\bf 49}, 35 (1999).
\bibitem{4} A.D. Sakharov, JETP Lett. {\bf 5}, 24 (1967).
\bibitem{5} A.I. Bochkarev, S.V. Kuzmin, and M.E. Shaposhnikov, Phys. Lett. B {\bf 244}, 275 (1990);
A.I. Bochkarev, S.V. Kuzmin, and M.E. Shaposhnikov, Mod. Phys. Lett. A {\bf 2}, 417 (1987).
\bibitem{6} M. Dine, R.G. Leigh, P. Huet, A. Linde, and D. Linde, Phys. Rev. D {\bf 46}, 550 (1992).
\bibitem{7} P. Arnold and O. Espinosa, Phys. Rev. D {\bf 47}, 3546 (1993);
Z. Fodor and A. Hebecker, Nucl. Phys. B {\bf 432}, 127 (1994).
\bibitem{8} K. Kajantie, M. Laine, K. Rummukainen, and M. Shaposhnikov, Phys. Rev. Lett. {\bf 77}, 2887 (1996); F. Csikor and Z. Fodor, and J. Heitger, Phys. Rev. Lett. {\bf 82}, 21 (1999).
\bibitem{9} G.W. Anderson and L.J. Hall, Phys. Rev. D {\bf 45}, 2685 (1992).
\bibitem{10} G.F. Giudice, Phys. Rev. D {\bf 45}, 3177 (1992).
\bibitem{11} J. Choi and R.R. Volkas, Phys. Lett. B {\bf 317}, 385 (1993);
J. Choi, Phys. Lett. B {\bf 345}, 253 (1995).
\bibitem{12} K. Enqvist, K. Kainulainen, and I. Vilja, Nucl. Phys. B {\bf 403}, 749 (1993);
I. Vilja, Phys. Lett. B {\bf 324}, 197 (1994).
\bibitem{13} A.I. Bochkarev, S.V. Kuzmin and M.E. Shaposhnikov, Phys. Rev. D {\bf 43}, 369 (1991);
N. Turok and J. Zadrozny, Nucl. Phys. B {\bf 358}, 471 (1991);
N. Turok and J. Zadrozny, Nucl. Phys. B {\bf 369}, 729 (1992);
L.D. McLerran, M.E. Shaposhnikov, N. Turok and M. Voloshin, Phys. Lett. B {\bf 256}, 451 (1991);
A.T. Davies, C.D. Froggatt, G. Jenkins, and, R.G. Moorhouse, Phys. Lett. B {\bf 336}, 464 (1994);
J.M. Cline and P.A. Lemieux, Phys. Rev. D {\bf 55}, 3873 (1997).
\bibitem{14} M. Carena, M. Quiros and C.E.M. Wagner, Phys. Lett. B {\bf 380}, 81 (1996);
M. Carena, M. Quiros and C.E.M. Wagner, Nucl. Phys. B {\bf 524}, 3 (1998);
B. de Carlos and J.R. Espinosa, Nucl. Phys. B {\bf 503}, 24 (1997);
M. Laine and K. Rummukainen, Phys. Rev. Lett. {\bf 80}, 5259 (1998);
M. Laine and K. Rummukainen, Nucl. Phys. B {\bf 535}, 423 (1998);
J.M. Cline and G.D. Moore, Phys. Rev. Lett.  {\bf 81}, 3315 (1998);
K. Funakubo, A. Kakuto, S. Otsuki, and F. Toyoda, Prog. Theor. Phys. {\bf 99}, 1045 (1998);
K. Funakubo, Prog. Theor. Phys. {\bf 101}, 415 (1999);
M. Losada Nucl. Phys. B {\bf 537}, 3 (1999).
\bibitem{15} M. Pietroni, Nucl. Phys. B {\bf 402}, 27 (1993);
A.T. Davies, C.D. Froggatt, R.G. Moorhouse, Phys. Lett. B {\bf 372}, 88 (1996);
S.J. Huber and M.G. Schmidt, Eur. Phys. J. C {\bf 10}, 473 (1999);
M. Bastero-Gil, C. Hugonie, S.F. King, D.P. Roy, S. Vempati, Phys. Lett. B {\bf 489}, 359 (2000).
\bibitem{16} A. Menon, D.E. Morrissey, and C.E.M. Wagner, Phys. Rev. D {\bf 70}, 035005 (2004);
S.W. Ham, S.K. Oh, C.M. Kim, E.J. Yoo, and D. Son, Phys. Rev. D {\bf 70}, 075001 (2004).
\bibitem{17} J. Kang, P. Langacker, T. Li, and T. Liu,  hep-ph/0402086.
\bibitem{18} W. Buchm\"uller and D. Wyler, Nucl. Phys. B {\bf 268}, 621 (1986);
C.N. Leung, S.T. Love, and S. Rao, Z. Phys. C {\bf 31}, 433 (1986);
\bibitem{19} K. Hagiwara, S. Ishihara, R. Szalapski, and D. Zeppenfeld, Phys. Lett. B {\bf 283}, 353 (1992);
K. Hagiwara, S. Ishihara, R. Szalapski, and D. Zeppenfeld, Phys. Rev. D {\bf 48}, 2182 (1993);
J. Wudka, Int. J. Mod. Phys. A {\bf 9}, 2301 (1994);
G.J. Gounaris, D.T. Papadamou, and F.M. Renard, Z. Phys. C {\bf 76}, 333 (1997);
T. Han, T. Huang, Z.H. Lin, J.X. Wang, and X. Zhang, Phys. Rev. D {\bf 61}, 015006 (2000);
Z.H. Lin, T. Han, T. Huang, J.X. Wang, and X. Zhang, Phys. Rev. D {\bf 65}, 014008 (2002);
V. Barger, T. Han, P. Langacker, B. McElrath, and P. Zerwas, Phys. Rev. D {\bf 67}, 115001 (2003).
\bibitem{20} A. Datta, B.L. Young, X. Zhang, Phys. Lett. B {\bf 385}, 225 (1996).
\bibitem{21} B. Grzadkowski and J. Wudka, Phys. Rev. Lett. {\bf 88}, 041802 (2002).
\bibitem{22} X. Zhang, Phys. Rev. D {\bf 47}, 3065 (1993).
\bibitem{23} C. Grojean, G. Servant, and J.D. Wells, hep-ph/0407019.
\bibitem{24} S. Coleman and E. Weinberg, Phys. Rev. D {\bf 7}, 1888 (1973).
\bibitem{25} M. Sher, Phys. Rep. {\bf 179}, 273 (1989).
\bibitem{26} L. Dolan and R. Jackiw, Phys. Rev. D {\bf 9}, 3320 (1974).

\end{thebibliography}
\end{document}